\documentclass[twocolumn,letter]{jpsj3}

\def\runauthor{Online-Journal Subcommittee}

\title{Relaxation Dynamics of Photocarriers in One-Dimensional Mott Insulators\\
Coupled to Phonons}
\author{H. Matsueda${}^{a}$, S. Sota${}^{b}$, T. Tohyama${}^{b}$, and S. Maekawa${}^{c,d}$}
\inst{
${}^{a}$Sendai National College of Technology, Sendai 989-3128, Japan \\
${}^{b}$Yukawa Institute for Theoretical Physics, Kyoto University, Kyoto 606-8502, Japan \\
${}^{c}$Advanced Science Research Center, Japan Atomic Energy Agency, Tokai 319-1195, Japan \\
${}^{d}$CREST, Japan Science and Technology Agency (JST), Sanbancho, Tokyo 102-0075, Japan}

\recdate{Received August 26, 2011; Accepted November 15, 2011; Published December 7, 2011} 

\abst{%

}

\kword{photoinduced phase transition, pump-probe reflection spectroscopy, strongly correlated electron systems, density matrix renormalization group method}

\begin{document}
\maketitle

It is a fundamental task in strongly correlated electron systems to examine interdependence among spin, charge, orbital, and lattice degrees of freedom. Recently, the examinations of the interdependence under nonequilibrium conditions become quite urgent issues in various on-going subjects. Ultrafast transient optics is an approach exploring new functionalities of materials and observing properties hidden in equilibrium conditions. The most important subject is photoinduced phase transition of low-dimensional transition metal oxides and organic materials with strong electron correlation. 

One-dimensional (1D) Mott insulators, Sr${}_{2}$CuO${}_{3}$ and halogen-bridged Ni compounds, show photoinduced insulator-to-metal transition accompanied with its picosecond recovery to the insulating state~\cite{Ogasawara,Iwai}. The time scale of the recovery is three order of magnitude faster than that for semiconductors~\cite{Nagai}. The 1D Mott insulators also exhibit gigantic third-order optical nonlinearity, and because of these two properties, they are promising future opto-electronics materials~\cite{Kishida}. The complete description of optically excited Mott insulators is thus desired, but key theoretical concepts in nonequilibrium are still under construction.

The photocarriers of the 1D Mott insulators are called holon and doublon representing empty and doubly occupied sites, respectively. A holon and a doublon recombine by emitting energy to other elementary excitations. A problem is to clarify a pass way of energy dissipation due to the recombination. Two possible candidates are spin and phonon excitations, since antiferromagnetic (AF) exchange energy and highest phonon frequencies are of the same order ($\sim$0.1~eV)~\cite{Suzuura}. Since high-energy states created by optical excitation may violate the separation of spin and charge degrees of freedom inherent in 1D electron systems~\cite{Takahashi}, a pass way for energy dissipation through a spin channel can be expected. However, recent numerical studies have shown robustness of the spin-charge separation for nonequilibrium steady states~\cite{Oka,Al-Hassanieh}. It is thus necessary to make clear a coupling of spin and charge degrees of freedom under photoirradiation. As for phonon relaxation, pump-probe experiments have been done for various TTF-TCNQ salts with different magnitudes of electron-phonon (EP) coupling~\cite{Okamoto}. K- and Rb-TCNQ show spin-Peierls (SP) transition at $T_{c}=$395~K and 220~K, respectively, and their photocarriers are once localized as polarons at around 70~fs, and then recombine with a few ps. On the other hand, ET-F${}_{2}$TCNQ does not show SP transition, and metallic photocarriers decay within 200~fs. Therefore, a fundamental question to be answered is about what is driving force of ultrafast relaxation of the 1D Mott insulators when the charge carriers couple weakly with spin and lattice. Since EP coupling is also present in semiconductors, we need to answer another question why phonon relaxation in the Mott insulators is much faster than that in the semiconductors.

In this Letter, we incorporate time dependent vector potential of laser pulse into density-matrix renormalization group (DMRG) simulation of a 1D Hubbard-Holstein model to answer the questions raised above. Examining initial relaxation after irradiation, we find that the spin-charge coupling exists for strong excitation in the case without phonon degrees of freedom. With introducing even small magnitude of EP coupling, the effect of the spin-charge coupling is suppressed, and many phonons are excited dynamically in the system. Phonon bases required for time-dependent calculation are much larger than those for the calculation of the optical conductivity with the same magnitude of EP coupling. This dynamical generation of phonons characterizes phonon relaxation in Mott insulators. We also discuss the difference of relaxation dynamics between Mott insulators and semiconductors.

We start with a 1D extended Hubbard-Holstein model with the classical vector potential of pump light:
\begin{eqnarray}
H(\tau) &=& -t\sum_{i,\sigma}(e^{iA(\tau)}c_{i,\sigma}^{\dagger}c_{i+1,\sigma}+{\rm H.c.}) \nonumber \\
&& + U\sum_{i}n_{i,\uparrow}n_{i,\downarrow} + V\sum_{i}(n_{i}-1)(n_{i+1}-1) \nonumber \\
&& + \omega_{0}\sum_{i}b_{i+1/2}^{\dagger}b_{i+1/2} \nonumber \\
&& - g\sum_{i}(b_{i+1/2}^{\dagger}+b_{i+1/2})(n_{i}-n_{i+1}),
\label{H}
\end{eqnarray}
where $c_{i,\sigma}^{\dagger}$ ($c_{i,\sigma}$) is a creation (annihilation) operator of an electron at site $i$ with spin $\sigma$, and $b_{i}^{\dagger}$ ($b_{i}$) is a creation (annihilation) operator of a phonon at site $i$. This model includes electron hopping, $t$, on-site and nearest-neighbor Coulomb repulsions, $U$ and $V$, respectively, phonon frequency, $\omega_{0}$, and EP coupling, $g$. The time-dependent vector potential, $A(\tau)$, is defined by
\begin{eqnarray}
A(\tau)=A_{0}e^{-(\tau-\tau_{0})^{2}/(2\tau_{d}^2)}\cos(\omega_\mathrm{pump}(\tau-\tau_{0})),
\label{A}
\end{eqnarray}
where $\tau_{d}$ is the duration time and $\omega_\mathrm{pump}$ is the frequency of pump laser. We note that, in the present study, both of electronic and phononic degrees of freedom are quantized in contrast to a recent work where phonons are treated as classical lattice vibrations~\cite{Yonemitsu}.

We calculate linear optical absorption spectrum before pumping, which is related to the current-current correlation function given by
\begin{equation}
\chi(\omega)=-\frac{1}{\pi}{\rm Im}\left<0\right|j\frac{1}{\omega+E_{0}-H+i\gamma}j\left|0\right>,
\end{equation}
with the current operator $j=it\sum_{i,\sigma}(c_{i,\sigma}^{\dagger}c_{i+1,\sigma}-{\rm H.c.})$ and the ground state $\left|0\right>$ with energy $E_0$.

Photodoping rate due to pumping, $\delta$, which is controlled by the amplitude of laser pulse $A_0$ in (\ref{A}), is defined by
\begin{equation}
\delta=\lim_{\tau\rightarrow\infty}(E(\tau)-E_{0})/(L\omega_\mathrm{pump}),
\end{equation}
where $L$ is the system size and $E(\tau)=\left<\tau\right|H\left|\tau\right>$. $\left|\tau\right>$ is the solution of the time-dependent Schr\"{o}dinger equation, $i(\partial/\partial\tau)\left|\tau\right>=H(\tau)\left|\tau\right>$, and the solution is given by
\begin{eqnarray}
\left|\tau\right>=T\exp\left\{-i\int_{0}^{\tau}H(\tau^{\prime})d\tau^{\prime}\right\}\left|0\right>
\end{eqnarray}
with $T$ being the time ordering operator.

We calculate the time evolution of doublon number, $N_\mathrm{d}(\tau)$, phonon number, $N_\mathrm{ph}(\tau)$, and local spin correlation, $C_\mathrm{s}(\tau)$. They are defined by
\begin{eqnarray}
N_\mathrm{d}(\tau)&=&\sum_{i}\left<\tau\right|n_{i,\uparrow}n_{i,\downarrow}\left|\tau\right>, \\
N_\mathrm{ph}(\tau)&=&\sum_{i}\left<\tau\right|b_{i}^{\dagger}b_{i}\left|\tau\right>,
\end{eqnarray}
and
\begin{eqnarray}
C_\mathrm{s}(\tau)&=&\sum_{i}\left<\tau\right|\vec{S}_{i}\cdot\vec{S}_{i+1}\left|\tau\right>.
\end{eqnarray}

For numerical calculation of these time-dependent quantities, we apply the DMRG technique as well as the numerically exact diagonalization method. In the DMRG technique, we perform a DMRG run for a given time $\tau$ and repeat this for many $\tau$. Here, we optimize just one time-dependent wave function $\left|\tau\right>$ for a given time, and thus can reduce the number of the target states in the reduced density matrix, leading to better performance. For the calcluation of $\left|\tau\right>$, we recursively use the following equations for small time slice $\Delta\tau$,
\begin{eqnarray}
\left|n\Delta\tau\right>&\simeq&\exp\left\{-iH(\tau_{n})\Delta\tau\right\}\left|(n-1)\Delta\tau\right> \\
&\simeq&\sum_{j=1}^{M}\left|j\right>e^{-iE_{j}\Delta\tau}\left<j\right|\left.(n-1)\Delta\tau\right>
\end{eqnarray}
until we obtain $\left|\tau\right>$. Here, $M$ represents the number of the Lanczos step, $\tau_{n}=(n-1/2)\Delta\tau$, $\left|j\right>$ and $E_{j}$ are the eigenvector and eigenenergy of the tri-diagonal matrix for $H(\tau_{n})$, respectively, and $\left<j\right|\left.(n-1)\Delta\tau\right>$ is the first component of $\left|j\right>$. Note that it is more sophisticated to do continuous time sweeping based on the adaptive time-dependent DMRG method~\cite{adaptive}. However, the wave function transformation crucial for this method only leads to poor numerical precision in the present model, and we do not use the adaptive method.

We consider half-filled systems, and the maximum system size is $L=12$. In our DMRG calculations, electron and phonon degrees of freedom are represented as different sites and an open boundary condition is used. The parameters $U$, $V$, and $\omega_{0}$ in (\ref{H}) are taken to be $10t$, $2t$, and $0.1t$, respectively~\cite{Matsueda}. Among parameters in (\ref{A}), $\tau_{d}$ and $\tau_{0}$ are fixed to be $\tau_{d}=2/t$ and $\tau_{0}=5/t$, respectively. The truncation number $m$ of density-matrix eigenvalues in the DMRG procedure is taken to be up to $m=1000$. The number of phonon is taken up to 6 states per phonon site.

\begin{figure}[tbp]
\begin{center}
\includegraphics[width=6cm]{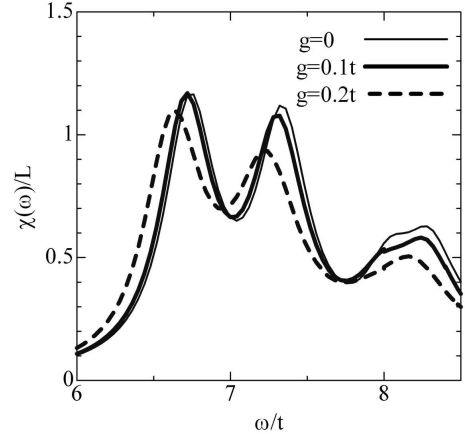}
\end{center}
\caption{Current-current correlation function $\chi(\omega)$ of a 1D extended Hubbard-Holstein chain for $L=12$ and a broadening factor $\gamma=0.2t$. The fine solid, bold solid, and bold dashed lines correspond to $g=0$, $0.1t$, and $0.2t$, respectively.}
\end{figure}

We first show $\chi(\omega)$ for various $g$ values in Fig.~1. The calculation was done by the dynamical DMRG and the obtained results agree qualitatively with previous numerical results~\cite{Matsueda}. Because of small system size, $\chi(\omega)$ exhibits discrete peaks with finite energy difference. For $g=0.1t$, line shape with a broadening factor $\gamma=0.2t$ is similar to that without $g$. In this case, it is enough to prepare two phonons on each site. In the following, we take $g=0.1t$ and $g=0.05t$ in order to make clear whether weak EP coupling affects the time evolution very strongly. Here, the effective EP coupling is defined by $\lambda=g^{2}/(2t\omega_{0})$, and the value of $\lambda$ is $\lambda=0.05$ and $\lambda=0.0125$ for $g=0.1t$ and $g=0.05t$, respectively.

\begin{figure}[tbp]
\begin{center}
\includegraphics[width=8cm]{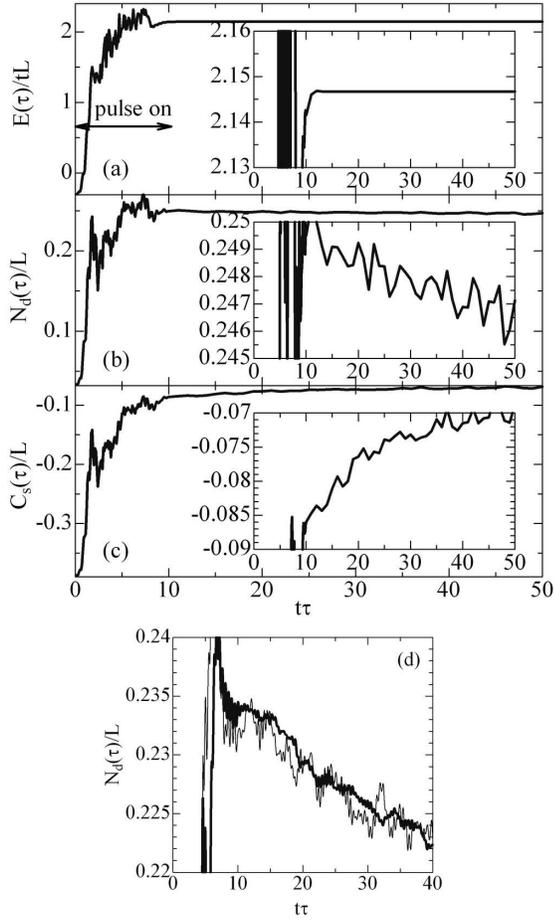}
\end{center}
\caption{(a-c) Time evolution of (a) $E(\tau)$, (b) $N_\mathrm{d}(\tau)$, and (c) $C_\mathrm{s}(\tau)$ in a $L=10$ extended Hubbard chain. Pump-laser irradiation continues up to $\tau\sim 10/t$ and photodoping rate $\delta=0.37$. Insets show enlarged scale of the corresponding quantities. Here, roughly 4 photons come in and then 2 doublons are created. Because of high density excitation, the number of photon and the number of doublon-holon pair are not the same. (d) System-size dependence of $N_\mathrm{d}(\tau)$ by exact diagonalization. Bold solid line for $L=10$ and fine solid line for $L=8$. We take $\omega_\mathrm{pump}=7t$ (close to the energy of the excitonic bound state) and $A_{0}=1.0$ ($\delta\sim 0.26$). 
}
\end{figure}

Before going to the extended Hubbard-Holstein model, we study the time evolution of the extended Hubbard model ($g=0$) to make clear how the coupling between photocarriers and spins is induced by strong laser pulse. In Fig.~2, we show time evolution of $E(\tau)$, $N_\mathrm{d}(\tau)$, and $C_\mathrm{s}(\tau)$. The photodoping rate $\delta$ is set to be $0.37$ corresponding to strong excitation. This case roughly corresponds to an experimental situation where clear Drude weight appears in absorption spectra just after pumping. $\omega_\mathrm{pump}$ is set to be $U-V=8t$, and then the optical pump mainly excite free carrier states. $E(\tau)$, $N_\mathrm{d}(\tau)$ and $C_\mathrm{s}(\tau)$ gradually increase during photoirradiation as expected. After the pumped pulse has been turned off ($\tau\ge 10/t=2\tau_{0}$), $E(\tau)$ is hardly changed because of treating a closed system. 

We find in Figs.~2(b) and 2(c) that $N_\mathrm{d}(\tau)$ decreases as $C_\mathrm{s}(\tau)$ increases with time (see insets) accompanied by oscillating behaviors. Unfortunately, the time resolution is limited in our DMRG due to the sparseness of time mesh. Thus, in order to see clearly the meaning of the decrease of $N_\mathrm{d}(\tau)$ and the oscillation, we show the exact diagonalization results of $N_\mathrm{d}(\tau)$ for $L=8$ and $L=10$ in Fig.~2(d). We take a pameter set different from that for Fig.~2(b), so that the oscillation looks clearer. Note that photoexcitation creates bound excitons in this case, and due to this fact the recombination rate is larger than that for Fig.~2(b). We see that the oscillation depends on the system size, while the decrease of $N_\mathrm{d}(\tau)$ does not. Therefore, we can safely say that the decrease of $N_\mathrm{d}(\tau)$ captures an essential feature of relaxation. The oscillation is composed of two periods. One is $\omega_\mathrm{pump}^{-1}$ which is much faster than our time mesh. The other one is roughly $L/t$ which can be observed in the DMRG data. Physically this time scale corresponds to electron motion from one edge of the chain to another edge. This is also related to descrete energy level structure of finite-size system.

We have confirmed that the gradual increase of $C_\mathrm{s}(\tau)$ seen in Fig.~2(c) is not clearly obse rved in cases of weak excitation~\cite{Takahashi}. The decrease of $N_\mathrm{d}(\tau)$ and the increase of $C_\mathrm{s}(\tau)$ mean that holon and doublon created by pump pulse recombine with each other and the recombination is accompanied by an energy flow from the photocarriers to the spin degree of freedom. However, it should be noted that the slope of time dependence in $C_\mathrm{s}(\tau)$ is very gradual. Furthermore, the change of $C_\mathrm{s}(\tau)$ from $t\tau=10$ to 50 ($\sim$0.2) is  smaller than the value of one spin flip ($\sim$1.0). Therefore, relaxation through the spin channel is not efficient in the extended Hubbard model, even though the spin-charge coupling becomes evident as the pump power increases. This is qualitatively consistent with recent reports showing robustness of the spin-charge separation for nonequilibrium steady states~\cite{Oka,Al-Hassanieh}.

\begin{figure}[tbp]
\begin{center}
\includegraphics[width=8cm]{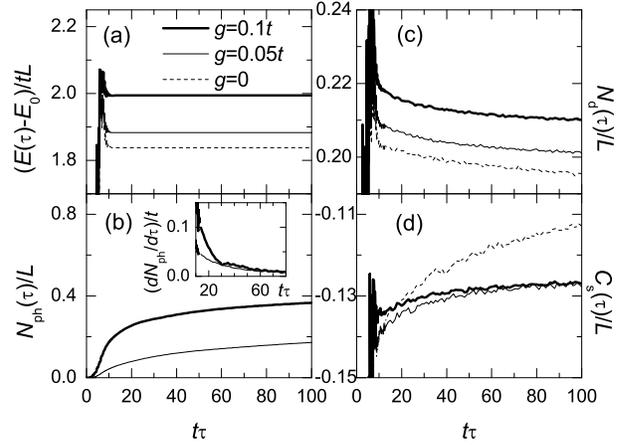}
\end{center}
\caption{Time evolution of various quantities in an extended Hubbard-Holstein chain with $L=12$. The bold solid, the fine solid, and the dashed lines represent the cases with $g=0.1t$, $g=0.05t$, and $g=0$, respectively. (a) $E(\tau)-E_0$, (b) $N_\mathrm{ph}(\tau)$, (c) $N_\mathrm{d}(\tau)$, and (d) $C_\mathrm{s}(\tau)$. The inset of (b) represents the first derivative of $N_\mathrm{ph}(\tau)$. We take $\omega_\mathrm{pump}=6.75t$ corresponding to the excitonic bound state (see Fig.~1), and $A_{0}=2.0$. Pump-laser irradiation continues up to $\tau\sim 10/t$.}
\end{figure}

The effect of EP coupling on relaxation dynamics is examined in Fig.~3, where $\omega_\mathrm{pump}$ is tuned to the excitonic bound states shown in Fig.~1. Until the pump pulse is tuned off, all of charge, spin, phonon degrees of freedom are disturbed. After pumping ($t\tau>10$), the doublon number $N_\mathrm{d}(\tau)$ decreases as is the case without EP coupling. In the latter case, the resulting energy loss in the charge degree of freedom was transferred to the spin channel accompanied by the decrease of spin correlation, i.e., increase of effective spin energy. In the former case with EP coupling, spin correlation is also reduced. However, the reduction of spin correlation is suppressed by increasing EP coupling, as is seen in Fig.~3(d). This implies the suppression of spin-charge coupling due to the presence of EP coupling.

The number of phonons increases after pumping as shown in Fig.~3(b). $N_\mathrm{ph}(\tau)$ does not contain oscillation, since the inverse phonon frequency $\omega_{0}^{-1}=10/t$ is slower than the other time scales.
We find for $g=0.1t$ that $N_\mathrm{ph}(\tau)$ in Fig.~3(b) has two regions in the time domain after pumping ($t\tau>10$): $N_\mathrm{ph}(\tau)$ increases rapidly and the slope changes at $t\tau\sim 30$ as is seen in the first derivative of $N_\mathrm{ph}(\tau)$ [the inset of Fig.~3(b)], followed by a gentle slope for $t\tau>30$. This characteristic time of kink $\tau\sim 30/t$ roughly corresponds to the duration time of laser pump ($\sim 10/t$) plus the inverse of effective EP coupling, $(\lambda t)^{-1}=20/t$. In a case with $g=0.05t$, the inverse EP coupling is $(\lambda t)^{-1}=80/t$, and the change in the slope is expected to occur at around $\tau\sim 90/t$. However, the kink structure is not clearly seen, since the coupling is too small. The rapid increase of phonon during $10/t<\tau<30/t$ indicates an energy transfer from charge degree of freedom to phonon one in the initial stage of relaxation. In other words, initial relaxation is dominated by phonons.

In the following, let us examine the mechanism of phonon relaxation in light of the presence of the two time regions. As shown in Fig.~3(b), the expectation value of phonon number per site is less than one, but the size of local phonon Hilbert space taken in the calculations is huge. Then, we have two possibilities for the relaxation processes. The first one is that the recombination of a holon and a doublon is acceralated by the phonon emission, if these carriers were still mobile against the EP coupling. In this case, the relaxation has been almost finished at around $t\tau=30$. On the other hand, if the phonon generation makes these carriers polaronic, the time scale $t\tau=30$ represents the polaron formation, eventually leading to slow recombination dynamics in the later time region. In the present case with $g=0.1t$, the former scenario is realized. We have tried preliminary calculation of a correlation function $\left<\tau\right|b_{j}^{\dagger}b_{j}n_{i,\uparrow}n_{i,\downarrow}\left|\tau\right>$ in a $6$-site chain, and have actually confirmed that the spatial distribution of phonons around photocarriers is uniform. When we take $g=0.2t$, the distribution starts to concentrate on the photocarrier site. The detailed results will be shown elsewhere.

Let us also estimate the typical value of the time $t\tau=30$. Our parameter set roughly corresponds to that for ET-F${}_{2}$TCNQ, since $g$ hardly affects the optical conductivity. The band width of the optical conductivity of our single-band model is $8t$, and experimentally the main band is located at around $0.5\sim 2.1$~eV. Then, the electron hopping is estimated to be $0.2$~eV. In this case, we obtain $\tau=30/t\sim 100$~fs. This value is not contradictory to the value $200$~fs for ET-F${}_{2}$TCNQ. 

In our previous study for the effect of $U$-dependence on charge-phonon coupling, we have found that photocarriers are dressed with phonon cloud strongly as $U$ increases for the large-$U$ region~\cite{Matsueda}. The result is quite suggestive for the difference of phonon relaxation dynamics between band and Mott insulators. In the large-$U$ limit, the model can be mapped onto a holon-doublon model. In this model, one holon-doublon pair exists and the spin degree of freedom is completely traced out. In the sense that only the charge degree of freedom remains, a phonon effect on these particles can be matched to that on charge carriers photo-doped into a band insulator. Taking this correspondence into account, we can judge that the photocarriers in the band insulator couple with phonons more strongly than in the Mott insulators. Strong EP coupling would accelerate localized polaron formation, leading to slow relaxation dynamics. The difference between band and Mott insulators is also seen in the DMRG calculation of spectral function~\cite{Matsueda2}, where we have needed a large number of local phononic states for band insulators in comparison with Mott insulators with the same EP coupling constant.

Summarizing, we have studied the relaxation dynamics of photocarriers in the 1D Mott insulators with EP coupling. The EP coupling dominates the spin-charge coupling in the initial relaxation, even if the optical conductivity is not affected by the EP coupling. We discussed the difference of relaxation between Mott and band insulators combining the present results with $U$-dependence on polaron formation.

This work was supported by the Next Generation Supercomputing Project of Nanoscience Program and Grant-in-Aid for Scientific Research from MEXT (19052003, 21740268, 22340097). A part of numerical calculations was performed in the supercomputing facilities in ISSP, University of Tokyo, YITP, Kyoto University, and IMR, Tohoku University. H. M. acknowledges hospitality of YKIS07 organized by the Yukawa International Program for Quark-Hadron Sciences at YITP. T.T. acknowledges the support from the Global COE Program "The Next Generation of Physics, Spun from Universality and Emergence".

\end{document}